\documentclass[prb,aps,superscriptaddress,twocolumn,showpacs,dvips]{revtex4}
\usepackage{feynmp}
\usepackage{amssymb}
\usepackage{amsmath}
\usepackage{epsfig}
\usepackage{array,graphicx,subfigure}
\usepackage{mathrsfs}
\usepackage{hyperref}
\usepackage{cases}
\usepackage{bbm}
\usepackage{bm}
\usepackage{gensymb}
\usepackage{float}
\usepackage{color}

\begin{document}

\title{Effects of random potentials in three-dimensional quantum electrodynamics}

\author{Peng-Lu Zhao}
\affiliation{Department of Modern Physics, University of Science and
Technology of China, Hefei, Anhui 230026, P. R. China}
\author{An-Min Wang}
\affiliation{Department of Modern Physics, University of Science and
Technology of China, Hefei, Anhui 230026, P. R. China}
\author{Guo-Zhu Liu}
\altaffiliation{gzliu@ustc.edu.cn} \affiliation{Department of Modern
Physics, University of Science and Technology of China, Hefei, Anhui
230026, P. R. China}

\begin{abstract}
Three-dimensional quantum electrodynamics exhibits a number of
interesting properties, such as dynamical chiral symmetry breaking,
weak confinement, and non-Fermi liquid behavior, and also has wide
applications in condensed matter physics. We study the effects of
random potentials, which exist in almost all realistic
condensed-matter systems, on the low-energy behaviors of massless
Dirac fermions by means of renormalization group method, and show
that the role of random mass is significantly enhanced by the gauge
interaction, whereas random scalar and vector potentials are
insusceptible to the gauge interaction at the one-loop order. The
static random potential breaks the Lorentz invariance, and as such
induces unusual renormalization of fermion velocity. We then
consider the case in which three types of random potentials coexist
in the system. The random scalar potential is found to play a
dominant role in the low-energy region, and drives the system to
undergo a quantum phase transition.
\end{abstract}


\maketitle

%%%%%%%%%%%%%%%%%%%%%%%%%%%%%Main Body%%%%%%%%%%%%%%%%%%%%%%%%%%%%%%%%%%%%%

\section{Introduction}

Massless three-dimensional quantum electrodynamics (QED$_{3}$)
describes the interaction between massless Dirac fermions and U(1)
gauge boson \cite{HeinzPRD81, AppelquistPRD81, Pisarski84}. This
field theory exhibits such non-perturbative phenomena as dynamical
chiral symmetry breaking (DCSB) \cite{Appelquist86, Appelquist88,
Nash89, Atkinson90, Maris96, Fisher04, Bashir08, Lo11, Feng13,
Feng14, Braun14} and weak confinement \cite{Burden92, Maris95}, and
thus is often regarded as a toy model of QCD. When the fermion
flavor is sufficiently large, the model is a conformal field theory
\cite{Klebanov12}. QED$_{3}$ and its variants have wide applications
in condensed matter physics: it is the low-energy effective theory
of high-$T_c$ cuprate superconductors \cite{Lee06, Affleck, Kim97,
Kim99, Rantner, Franz, Herbut, Liu02, Liu03} and certain spin liquid
systems \cite{Ran07, Hermele08, Lu14, Metlitski15, Wang15, Mross15}.
The non-perturbative phenomenon of DCSB provides an elegant
field-theoretic description of the two-dimensional Heisenberg
quantum antiferromagnetism \cite{Affleck, Kim99, Franz, Herbut,
Liu02, Liu03}, whereas the non-Fermi liquid behaviors induced by
gauge interaction may be used to understand the observed unusual
normal state of high-$T_c$ superconductors \cite{Lee06, Kim97,
Rantner, Franz, Herbut, WangLiu10A, WangLiu10B}. For these reasons,
QED$_{3}$ has attracted considerable research interest in the
communities of both high energy and condensed matter physics.

Previous works studying QED$_{3}$ have mainly focused on DCSB
\cite{Appelquist86, Appelquist88, Nash89, Atkinson90, Maris96,
Fisher04, Bashir08, Lo11, Feng13, Feng14, Braun14} and non-Fermi
liquid behaviors \cite{WangLiu10A, WangLiu10B, WangLiu12} caused by
the U(1) gauge boson in the clean limit. The effects of random
potential are rarely considered in the literature. In a realistic
condensed-matter system, there are always certain amount and types
of random potential, which may substantially affect the dynamics of
massless Dirac fermions. If some random potential is a relevant
perturbation to the system, it can determine many of the low-$T$
transport properties of Dirac fermions. Moreover, random potential
can also lead to instabilities of the system, which would drive
various kinds of quantum phase transition. To broaden the
applicability of QED$_{3}$ in condensed-matter physics, it is
necessary to examine the impact of various types of random
potential.

In this work, we analyze the roles played by random potentials in
the low-energy region and determine all the possible infrared fixed
points. To make a unbiased analysis, we shall treat the gauge
interaction and random potential equally, and study their interplay
by means of renormalization group (RG) method. Depending on the
value of fermion flavor $N$, QED$_{3}$ stays in the DCSB phase for
small $N$ and chirally symmetric phase for large $N$. Here, we
suppose a large $N$ and keep Dirac fermions massless. The random
potential is assumed to be static, and might be caused by defects
and/or impurity atoms in various realistic condensed-matter systems.
Generically, there are three types of random potential that can
couple to Dirac fermions \cite{Ludwig94, Nersesyan95, Altland02,
Stauber2005PRB, HerbutPRL2008, Vafek08, Wang2011PRB, WangLiu14}:
random mass (RM), random scalar potential (RSP), and random vector
(gauge) potential (RVP). We will first study the impact of each
single random potential, and then examine how different types of
random potential affect each other.

RG analysis show that the random potentials can lead to unusual
renormalization of fermion velocity $v_F$ as a consequence of
explicit Lorentz symmetry breaking. The role played by RM can be
significantly enhanced by the gauge interaction, but the roles
played by RVP and RSP are nearly unchanged by the gauge interaction.
We also study the fixed point structure of the system with all three
types of random potential present simultaneously. In this case, RSP
is much more important than RM and RVP in the low-energy region, and
derive the system to undergo a diffusive quantum phase transition,
which occurs even when RSP is quite weak. In the absence of RSP, we
find that RVP promotes the role of RM and also induce an anomalous
dimension for $v_F$.

The rest of the paper is organized as follows. We present the whole
action and the corresponding Feynman rules in Sec.~\ref{Sec_action},
and derive the RG equations in Sec.~\ref{Sec_Leading_RG}. The impact
of each single type of random potential and the mutual influence
between different random potentials are analyzed in
Sec.~\ref{Sec_single_disorder}. We summarize the results and
highlight possible future works in Sec.~\ref{Sec_summary}.

\section{Effective action}\label{Sec_action}

The Lagrangian density of QED$_{3}$ with $N$ flavors of massless
Dirac fermions is given by
\begin{eqnarray}
\mathcal{L}_{F} = \sum_{\sigma=1}^{N}\bar{\psi}_{\sigma}
\gamma^{\mu}(\partial_{\mu}+ie A_{\mu})\psi_{\sigma} -
\frac{1}{4}F_{\mu\nu}^{2},\label{Eq:Langrangian}
\end{eqnarray}
where $\partial_{\mu}= (\partial_0, v_F \partial_i)$ with $i=1,2$.
The electromagnetic tensor is $F_{\mu\nu} = \partial_{\mu}A_{\nu} -
\partial_{\nu}A_{\mu}$. Here, the Dirac fermion is described by a
four-component spinor $\psi$, whose conjugate is defined as
$\bar{\psi} = \psi^{\dag}\gamma_{0}$. The gamma matrices can be
chosen as $(\gamma_{0},\gamma_{1},\gamma_{2}) =
(\sigma_{3},\sigma_{2},-\sigma_{1})\otimes\sigma_{3}$, which satisfy
the Clifford algebra $\{\gamma_{\mu},\gamma_{\nu}\} =
2\delta_{\mu\nu}$. In (2+1) dimensions, there are two chiral
matrices, denoted by $\gamma_{3}=I_{2\times2}\otimes\sigma_{1}$ and
$\gamma_{5} = -I_{2\times2}\otimes\sigma_{2}$ respectively, which
anti-commute with $\gamma_{0,1,2}$. The model contains two
parameters: electric charge $e$, and fermion velocity $v_F$. It is
easy to check that $e$ is dimensional, so this field theory is
renormalizable and thus safe in the UV region. However, the gauge
interaction becomes strong in the IR region, which might cause
nontrivial physics.

The full theory respects the Lorentz symmetry, the U(1) local gauge
symmetry, and an additional continuous $U(2N)$ chiral symmetry $\psi
\rightarrow e^{i\theta \gamma_{3,5}}\psi$ with $\theta$ being an
arbitrary constant if the fermions are massless. The local gauge
symmetry is robust, but the other two symmetries can be easily
broken, either explicitly or dynamically.

Extensive previous studies \cite{Appelquist86, Appelquist88, Nash89,
Maris96, Fisher04, Bashir08} have confirmed that a finite fermion
mass can be dynamically generated by the gauge interaction if the
fermion flavor $N$ is smaller than certain critical value $N_c$,
which leads to DCSB. In the DCSB phase, the massive fermions are
confined by a logarithmic potential \cite{Burden92, Maris95}. For $N
> N_c$, the Dirac fermions remain massless and thus the theory
preserves the chiral symmetry. In the chirally symmetric phase, the
physical properties are far from trivial as the strong gauge
interaction can induce non-Fermi liquid behaviors of Dirac fermions
\cite{Lee06, Kim97, Rantner, Franz, Herbut, WangLiu10A, WangLiu10B}.

At zero temperature, the Lorentz invariance is strictly preserved.
In this case, the fermion velocity $v_F$ does not renormalize at all
and remains a constant. If the Lorentz invariance is broken, the
gauge interaction would renormalize $v_F$, which then exhibits
explicit dependence on momenta and energy. Thermal fluctuation
definitely breaks the Lorentz invariance, and hence leads to
velocity renormalization \cite{WangLiu2016}. If we stay at zero
temperature but includes static random potential, the Lorentz
invariance is also explicitly broken. As a result, the fermion
velocity will be renormalized.

We now incorporate random potential into the Lagrangian density of
QED$_3$ by writing down the following term \cite{Ludwig94,
Nersesyan95, Altland02, Stauber2005PRB, HerbutPRL2008, Vafek08,
Wang2011PRB, WangLiu14},
\begin{eqnarray}
\mathcal{L}_{d}=\sum_{\sigma=1}^{N}\bar{\psi}_{\sigma}\left(\sum_{\Gamma}
V_{\Gamma}(\mathbf{x})\Gamma\right)\psi_{\sigma}
,\label{Eq:Langrangian_d}
\end{eqnarray}
where the function $V_{\Gamma}(\mathbf{x})$ stands for the randomly
distributed potential. We assume $V_{\Gamma}(\mathbf{x})$ to be a
quenched, Gaussian white noise potential characterized by the
following identities:
\begin{eqnarray}
\langle V_{\Gamma}(\mathbf{x})\rangle = 0,\qquad \langle
V_{\Gamma}(\mathbf{x})V_{\Gamma}(\mathbf{x}')\rangle =
\Delta_{\Gamma} \delta^2(\mathbf{x}-\mathbf{x}').\label{Eq_def_dis}
\end{eqnarray}
The random potential is classified by the expression of matrix
$\Gamma$: $\Gamma=\mathbb{I}_{4\times 4}$ for RM; $\Gamma=\gamma_0$
for RSP; $\Gamma=(i\gamma_1,i\gamma_2)$ for RVP. It is also possible
to include other types of random potential, but these three types
are most frequently studied. The random potential can be induced by
various mechanisms in realistic Dirac fermion materials~\cite{Nersesyan95, CastroNeto,Peres2010RMP,Mucciolo2010JPCM,Meyer2007Nature,Champel2010PRB,Kusminskiy2011PRB}. These three types of random potential might
exist individually, or coexist in the same material. We will first
consider the impact of each single random potential, and then study
their mutual influence.

The random potential $V(\mathbf{x})$ needs to be properly averaged.
The simplest and most widely used scheme is to average over
$V(\mathbf{x})$ by employing the replica method \cite{LeeRMP1985,
Lerner, Edwards1975, Goswami2011PRL, LaiarXiv, Roy2014PRB,
Roy2016PRB, Roy2016SCP}, which leads us to an effective replicated
action written in the Euclidean space:
\begin{eqnarray}
\overline{S} &=& \int d^2x d\tau \Big\{\bar{\psi}_{\sigma}^{\alpha}
\left[\gamma_0(\partial_0+ieA_0) + \gamma_j(v_{F}\partial_j +
ieA_j)\right]\psi_{\sigma}^{\alpha}
\nonumber \\
&&-\frac{1}{4}F_{\mu\nu}^{2}\Big\} - \frac{1}{2}\int d^2x d\tau
d\tau' \Big[\Delta_{M}\big(\bar{\psi}_{\sigma}^{\alpha}
\psi_{\sigma}^{\alpha}\big)_x \big(\bar{\psi}_{\sigma}^{\beta}
\psi_{\sigma}^{\beta}\big)_{x'}\nonumber \\
&&+\Delta_S\big(\bar{\psi}_{\sigma}^{\alpha}\gamma_0
\psi_{\sigma}^{\alpha}\big)_x
\big(\bar{\psi}_{\sigma}^{\beta}\gamma_0
\psi_{\sigma}^{\beta}\big)_{x'}
+\Delta_V\big(\bar{\psi}_{\sigma}^{\alpha}i\gamma_j
\psi_{\sigma}^{\alpha}\big)_x \nonumber \\
&&\times\big(\bar{\psi}_{\sigma}^{\beta}i\gamma_j
\psi_{\sigma}^{\beta}\big)_{x'}\Big]\label{Eq_action}.
\end{eqnarray}
Here, $\alpha$ and $\beta$ are the replica indices, and $x
\equiv(\mathbf{x},\tau)$ and $x'\equiv(\mathbf{x},\tau')$. All the
repeated indices are summed up automatically. To distinguish
different types of random potential, we have introduced three new
parameters $\Delta_M$, $\Delta_S$, and $\Delta_V$ to characterize
the effective strength of quartic couplings of Dirac fermions
induced by averaging over RM, RVP, and RSP, respectively.

We choose to work in the Euclidean space, and write the free fermion
propagator as
\begin{eqnarray}
G_0(k_0,\mathbf{k}) = \frac{-i}{\gamma_0 k_0 + v_F\bm{\gamma}\cdot
\mathbf{k}}.
\end{eqnarray}
The free gauge boson propagator under Landau gauge reads
\begin{eqnarray}
D^0_{\mu\nu}(q) = \frac{v_F^2}{Q^2}\left(\delta_{\mu\nu} -
\frac{Q_{\mu}Q_{\nu}}{Q^2}\right),\label{Eq:gauge_pro}
\end{eqnarray}
where $Q_{\mu}\equiv(q_0,v_F\mathbf{q})$ and $Q^2 = Q_{\mu}Q_{\mu} =
q_{0}^2 + v_F^2\mathbf{q}^2$.

In the next section, we will perform RG calculations starting from
Eq.~(\ref{Eq_action}). The interaction between Dirac fermions and
gauge boson is treated by making a $1/N$ expansion by supposing a
general large $N$. For large value of $N$, DCSB cannot take place
and the Dirac fermions are kept massless throughout our
calculations. However, the parameters $\Delta_M$, $\Delta_S$, and
$\Delta_V$ are assumed to be small, corresponding to the nearly
clean case.

\section{Derivation of RG equations}\label{Sec_Leading_RG}

In this section, we calculate the quantum corrections to the
polarization tensor, fermion self-energy, fermion-disorder vertex,
and gauge coupling vertex to the leading order of perturbative
expansion. Based on these results, we will be able to derive the RG
flow equations for all the free model parameters.

\subsection{Polarization tensor and fermion self-energy}

\begin{figure}[htbp]
\center
\includegraphics[width=1.7in]{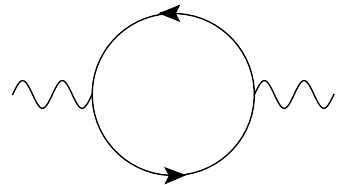}
\vspace{-0.20cm} \caption{One-loop Feynman diagram for polarization
tensor, where solid line stands for the free fermion propagator and
wavy line for the bare gauge boson propagator.}
\label{Fig_polarization}
\end{figure}

At the one-loop level, the diagram for the polarization function are
shown in Fig.~\ref{Fig_polarization}. It is straightforward to get
\begin{eqnarray}
\Pi_{\mu\nu}(q) &=& Ne^2\int\frac{d^{3}k}{(2\pi)^{3}} \mathrm{Tr}
\left[G(k)\gamma_{\mu}G(k+q)\gamma_{\nu}\right], \nonumber \\
&=& Ne^2 \frac{1}{v_F^2}\int\frac{d^{3}K}{(2\pi)^{3}} \mathrm{Tr}
\left[G(K)\gamma_{\mu}G(K+Q)\gamma_{\nu}\right],\nonumber \\
&=& \Pi(Q)\left(\delta_{\mu\nu}-\frac{Q_{\mu}Q_{\nu}}{Q^2}\right),
\end{eqnarray}
where the function
\begin{eqnarray}
\Pi(Q) = -\frac{\alpha Q}{v_F^2}
\end{eqnarray}
with $\alpha = Ne^2/8$. For massless QED$_3$, the dimensional
coupling $\alpha = Ne^2/8$ is kept fixed as $N\rightarrow \infty$,
providing the only fixed energy scale in the theory
\cite{AppelquistPRD81, HeinzPRD81, Pisarski84, Appelquist86,
Appelquist88, Nash89}. Including the corrections to the polarization
function, we write the effective gauge boson propagator in the form
\begin{eqnarray}
D_{\mu\nu}(Q) &=& \frac{v_F^2}{Q^2 + \alpha Q}\left(\delta_{\mu\nu}
- \frac{Q_{\mu}Q_{\nu}}{Q^2}\right) \nonumber\\
&\approx&\frac{v_F^2}{\alpha Q}\left(\delta_{\mu\nu} -
\frac{Q_{\mu}Q_{\nu}}{Q^2}\right).\label{Eq.gauge_pro}
\end{eqnarray}
The quantum corrections is rapidly damped for momenta $Q > \alpha$
\cite{Appelquist88, Nash89, Maris96}, thus the above approximation
is well justified and also has been widely used \cite{Kim97, Kim99,
Franz}.

\begin{figure}[htbp]
\center \hspace{-2ex} \subfigure{
\includegraphics[width=1.6in]{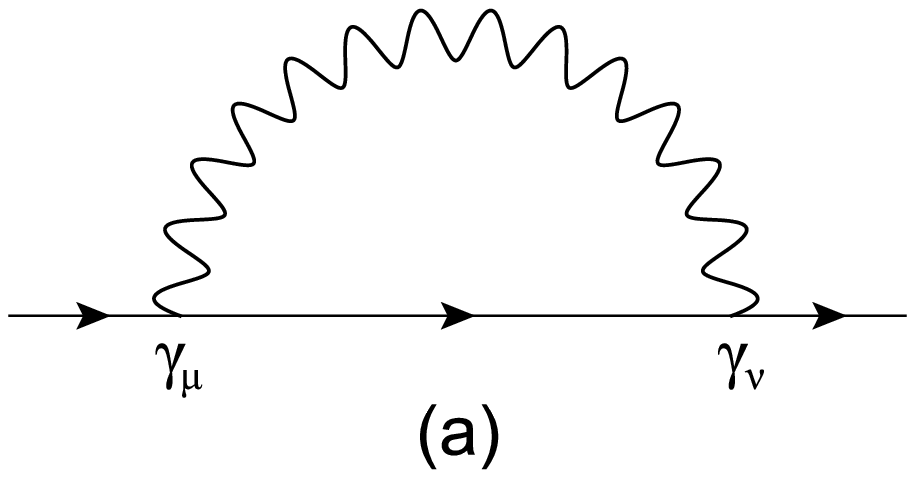}\label{Fig_Gauge_self_energy}}
\subfigure{
\includegraphics[width=1.6in]{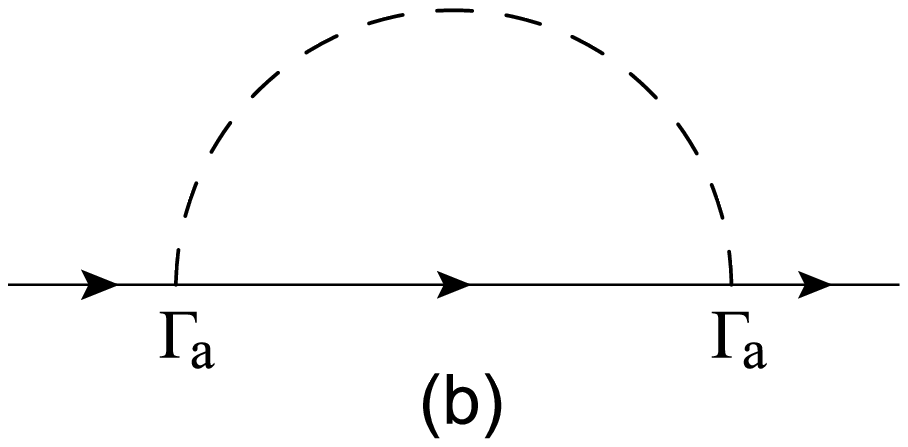}\label{Fig_Disorder_self_energy}}
\caption{One-loop fermion self-energy due to (a) gauge interaction;
(b) random potential (dashed line).} \label{Fig_Fermion_SE}
\end{figure}

Diagrams for fermion self-energy are shown in
Fig.~\ref{Fig_Fermion_SE}. According to
Fig.~\ref{Fig_Gauge_self_energy}, the correction due to gauge
interaction to the leading order of $1/N$ expansion is
\begin{eqnarray}
\Sigma^{G}(p_{0},\mathbf{p}) &=& -e^2\int \frac{d^3q}{(2\pi)^3}
\gamma_{\mu} G(p-q)\gamma_{\nu} D_{\mu\nu}(Q)\nonumber \\
&=& i\eta_{\psi}\gamma_{\mu}P_{\mu}\ln b. \label{Eq.Self_energy_G}
\end{eqnarray}
Here, the momenta integration is restricted within the shell
$Q\in[\Lambda/b,\Lambda]$, where $\Lambda$ is an UV cutoff and $b =
e^{l}$ with $l\geq$ being a freely varying length scale. We use
$\eta_{\psi}$ to denote the anomalous dimension of the fermion wave
function renormalization. To the leading order, we have
\begin{eqnarray}
\eta_{\psi} = \frac{8}{3\pi^2 N},
\end{eqnarray}
which is in accordance with Refs.~\cite{Atkinson90, Maris96}.


Form the above calculations, we can see that the fermion velocity
$v_F$ is not renormalized at all, which means $v_F$ is a constant
independent of varying energy scale. It is therefore safe to set
$v_F \equiv 1$ and recover $v_F$ whenever necessary. However, when
static random potential is added to the system, the Lorentz symmetry
is explicitly broken, and $v_F$ may received singular corrections.

According to Fig.~\ref{Fig_Disorder_self_energy}, the one-loop
disorder-induced fermion self-energy is given by
\begin{eqnarray}
\Sigma_{\mathrm{dis}}(k_0) &=& \sum_{\Gamma}\Delta_{\Gamma}
\int\frac{d^2\mathbf{k}}{(2\pi)^2}\Gamma
G_{0}(k_0,\mathbf{k})\Gamma \nonumber \\
&=&-ik_0\sum_{\Gamma}\frac{\Delta_{\Gamma}
\Gamma\gamma_0\Gamma}{2\pi v_{F}^{2}}\ln b \nonumber \\
&=&-ik_0\gamma_0\frac{\Delta_S+\Delta_{M} + 2\Delta_V}{2\pi
v_{F}^{2}}\ln b. \label{Eq.Self_energy_d}
\end{eqnarray}
It is clear that random potential does not lead to wave function
renormalization of spatial components, so $v_{F}$ will be
renormalized.

\subsection{Gauge coupling and fermion-disorder vertex}

Besides the gauge coupling parameter, disorder also bring another
kind of parameter which is the effective strength of coupling
between fermion and disorder. At this subsection, these vertices
corrections are computed.

\begin{figure}[htbp]
\center \hspace{-2ex} \subfigure{
\includegraphics[width=1.55in]{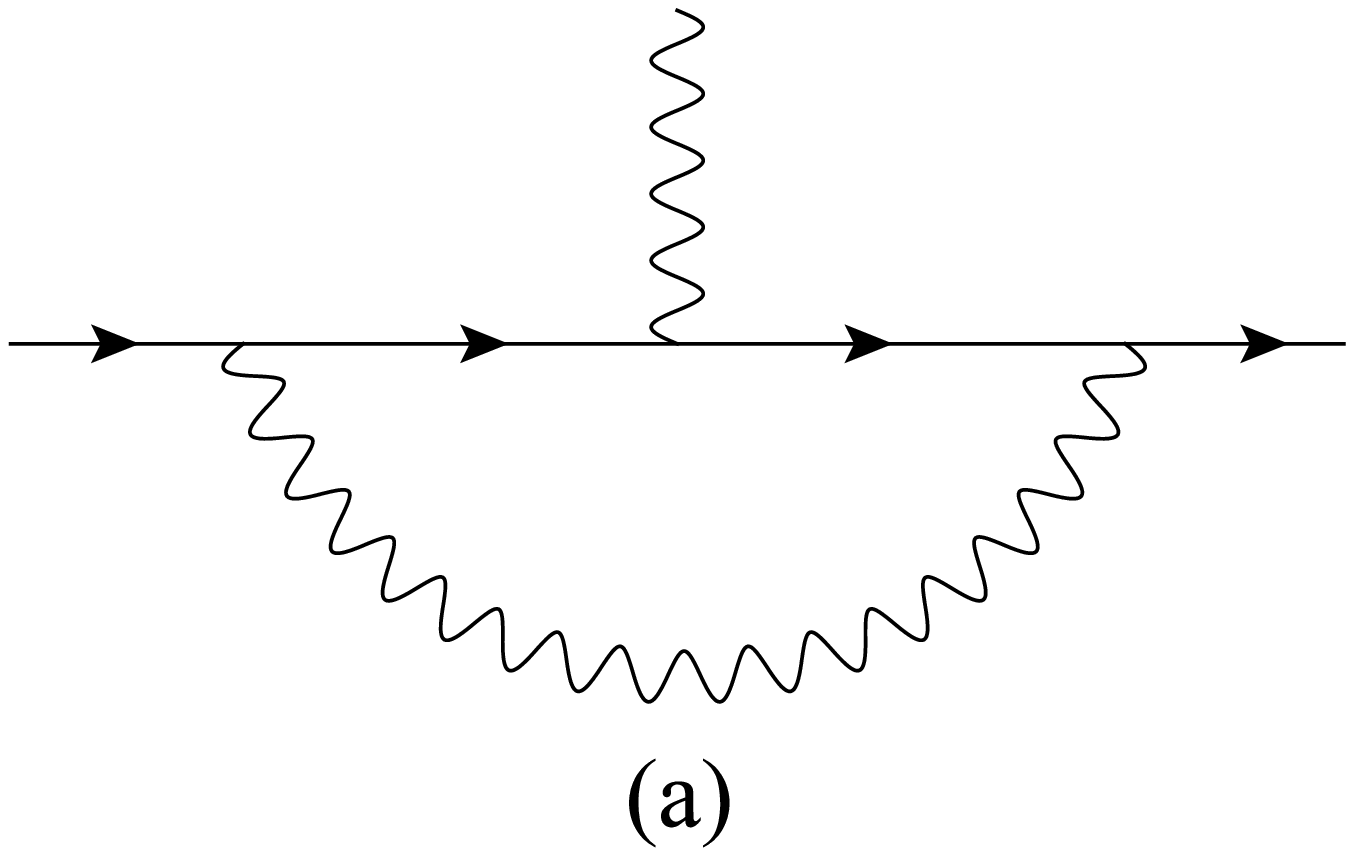}
\label{Fig_Gauge_Vertex_G}}
\subfigure{\includegraphics[width=1.55in]{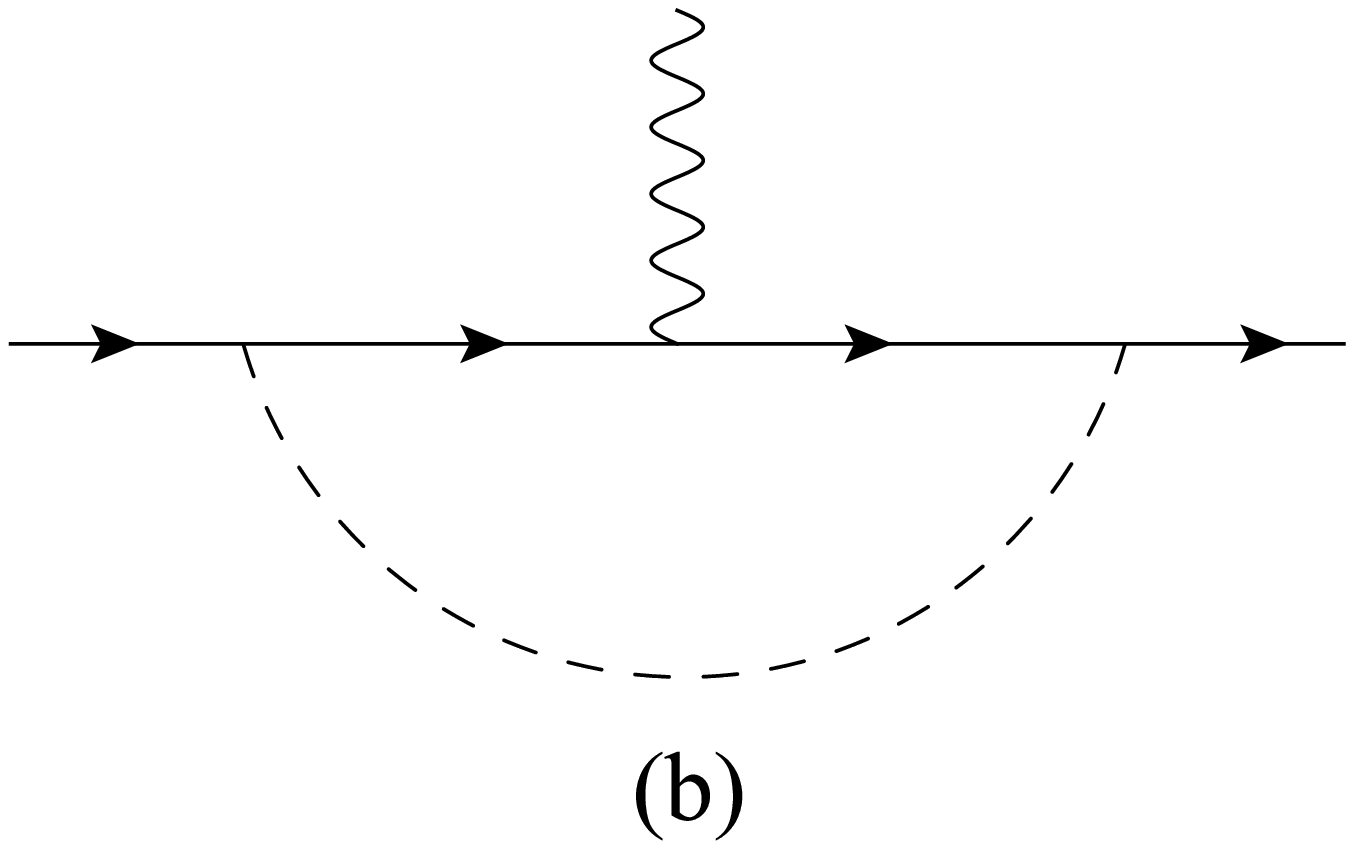}
\label{Fig_Gauge_Vertex_d}} \caption{One-loop gauge coupling
correction due to (a) gauge interaction, (b) random potential.}
\label{Fig_Gauge_Vertex}
\end{figure}

The one-loop diagrams gauge coupling corrections are depicted in
Fig.~\ref{Fig_Gauge_Vertex}. At vanishing external momenta and
energy, the vertex correction due to gauge interaction shown in
Fig.~\ref{Fig_Gauge_Vertex_G} is
\begin{eqnarray}
V_{\mathrm{e}}^{G} &=& -ie^3\int\frac{d^3q}{(2\pi)^3}
\gamma_{\rho}G_{0}(q)\gamma_{\mu} G_{0}(q)\gamma_{\nu}D_{\rho\nu}(Q)
\nonumber \\
&=& -ie\gamma_{\mu}\eta_{\psi}\ln b. \label{Eq.Gauge_Vertex_G}
\end{eqnarray}
According to Fig.~\ref{Fig_Gauge_Vertex_d}, the gauge coupling
correction due to random potential at zero external momenta-energy
is
\begin{eqnarray}
V_{\mathrm{e}}^{d}&=&ie \sum_{\Gamma}\Delta_{\Gamma}
\int\frac{d^2\mathbf{k}}{(2\pi)^2}\Gamma
G_{0}(k_0,\mathbf{k})\gamma_{\mu}G_{0}(k_0,\mathbf{k})\Gamma \nonumber \\
&=&ie\gamma_0\frac{\Delta_S+\Delta_{M}+2\Delta_V}{2\pi v_{F}^{2}}\ln
b. \label{Eq.Gauge_Vertex_d}
\end{eqnarray}

\begin{figure}[htbp]
\center
\subfigure{\includegraphics[width=3.2in]{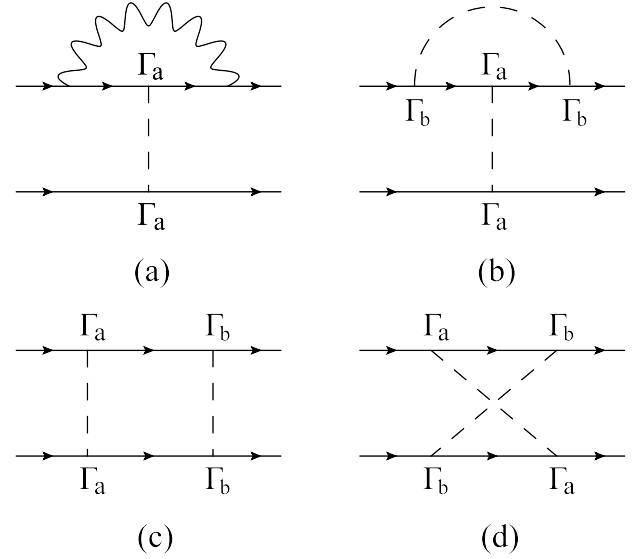}}
\caption{One-loop corrections to the coupling vertex between
fermions and random potentials in the vanishing replica
limit.}\label{Fig_all_disorder_vertex}
\end{figure}

In the replica limit, the one loop Feynman diagrams for the
corrections to fermion-disorder vertex are shown in
Fig.~\ref{Fig_all_disorder_vertex}. At zero external momentum and
frequency, the corresponding correction induced by gauge interaction
depicted in Fig.~\ref{Fig_all_disorder_vertex}(a) is calculated as
\begin{eqnarray}
V_{\mathrm{dis}}^{G}&=&e^2\Delta_{\Gamma}\int\frac{d^3q}{(2\pi)^3}
\gamma_{\mu}G_{0}(q)\Gamma G_{0}(q)\gamma_{\nu}D_{\mu\nu}(Q)\nonumber\\
&=&a\eta_{\psi}\ln b(\Delta_{\Gamma}\Gamma),\label{Eq.Disorder_Vertex_G}
\end{eqnarray}
where $a=-3$ for RM, and $a=1$ for RSP and RVP.
Fig.~\ref{Fig_all_disorder_vertex}(b) is the vertex correction due
to disorder averaging, and given by
\begin{eqnarray}
V_{\mathrm{dis}}^{d} &=& \Delta_a \sum_{\Gamma_b}\Delta_b
\int\frac{d^2\mathbf{q}}{(2\pi)^2}\Gamma_b G_0(q_0,\mathbf{q})
\Gamma_a G_0(q_0,\mathbf{q})\Gamma_b,\nonumber \\
\label{Eq.Disorder_Vertex_d}
\end{eqnarray}
After analytical calculations, we get
\begin{eqnarray}
V_{\mathrm{dis}}^{d} &=& \frac{-(\Delta_S+\Delta_{M} +
2\Delta_V)\Delta_{M}}{2\pi v_{F}^{2}}\mathbb{I}_{4\times4}\ln b
\label{Eq.Disorder_Vertex_dI}
\end{eqnarray}
for RM,
\begin{eqnarray}
V_{\mathrm{dis}}^{d} &=& \frac{(\Delta_S+\Delta_{M}-
2\Delta_V)\Delta_S}{2\pi v_{F}^{2}}\gamma_0\ln b,
\label{Eq.Disorder_Vertex_d0}
\end{eqnarray}
for RSP, and
\begin{eqnarray}
V_{\mathrm{dis}}^{d}&=&
0. \label{Eq.Disorder_Vertex_d1}
\end{eqnarray}
for RVP. The sum of the two diagrams given by
Fig.~\ref{Fig_all_disorder_vertex}(c) and
Fig.~\ref{Fig_all_disorder_vertex}(d) produces a nonzero correction
to another type of random potential defined by the matrix $\Gamma =
\gamma_0\bm{\gamma}$ along with parameters $(\Gamma_a,\Gamma_b) =
(\mathbb{I}_{4\times4},\gamma_0)$ \cite{Goswami2011PRL,Roy2016SCP}.
However, this type of random potential is not considered in the
present paper. For the three types of random potential under
consideration, the contributions from
Fig.~\ref{Fig_all_disorder_vertex}(c) and
Fig.~\ref{Fig_all_disorder_vertex}(d) simply cancel each other by
virtue of the relation \cite{Roy2014PRB, LaiarXiv}: $G(-k_0,
-\mathbf{k}) = -G(k_0,\mathbf{k})$.

\begin{figure*}[htbp]
\center \hspace{-4ex}\subfigure{\begin{minipage}{7.8cm}
\includegraphics[width=2.9in]{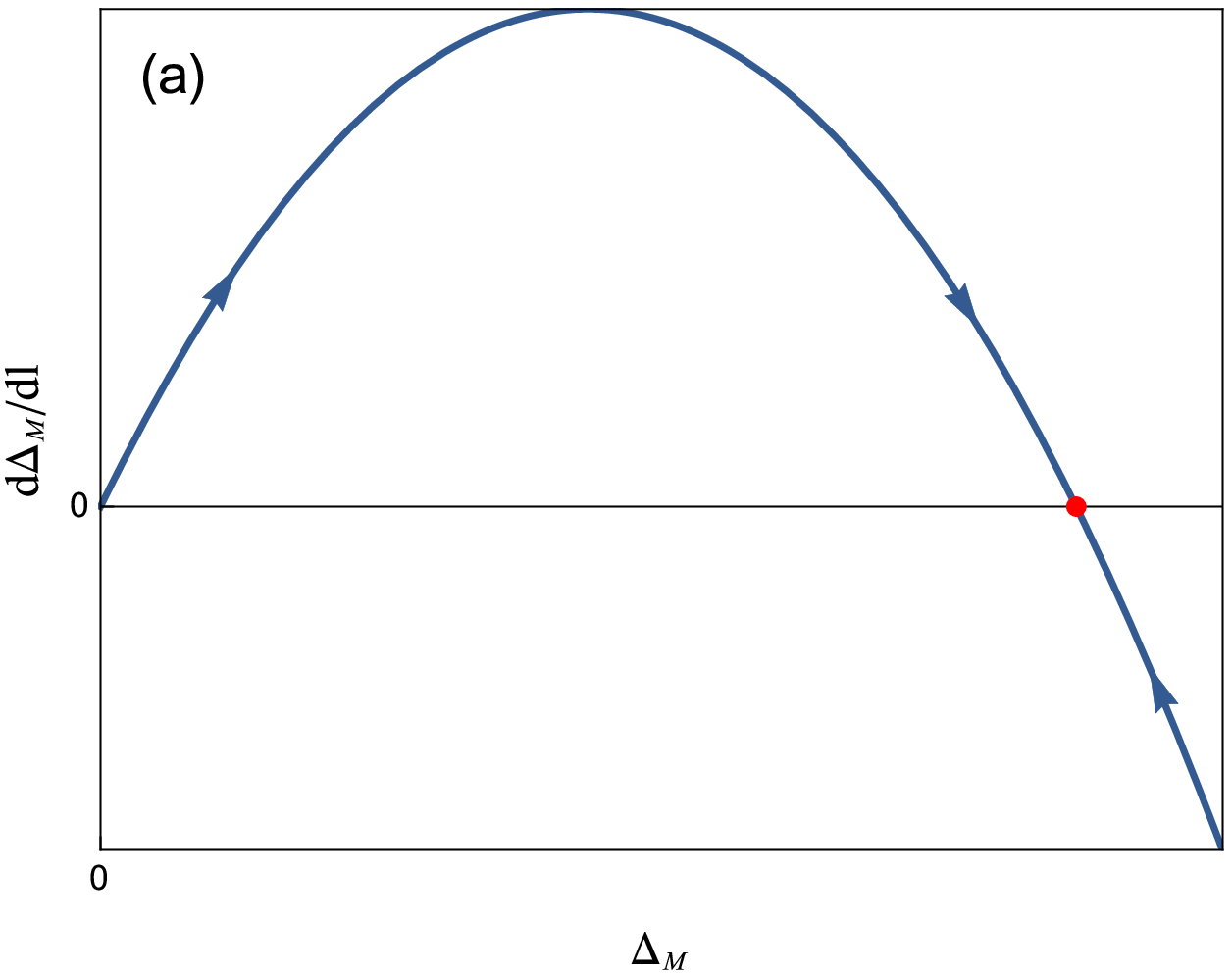}\label{Fig_random_mass_a}
\vspace{-1.3ex}
\end{minipage}}\hspace{2ex}
\subfigure{\begin{minipage}{7.8cm}
\includegraphics[width=2.93in]{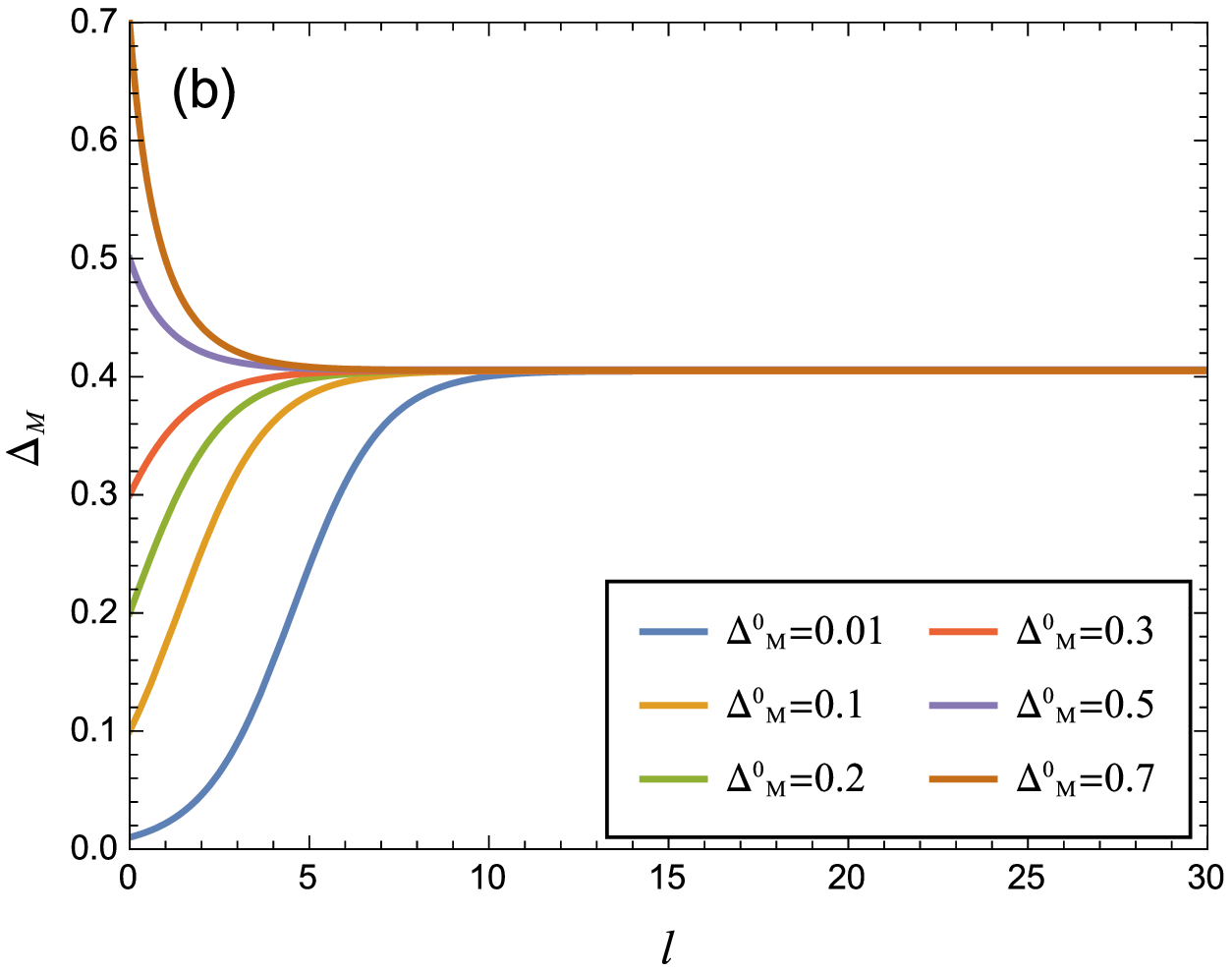}\label{Fig_random_mass_b}
\end{minipage}}\caption{(a) Flow diagram of $\Delta_M$
for RM. There is an unstable Gaussian fixed point $\Delta_M = 0$ and
a finite stable fixed point $\Delta_M = 4\eta_{\psi} + 2\Delta_V^0$.
(b) Dependence of $\Delta_M$ on the running scale $l$ at different
initial values. Here, $N = 4$ and $\Delta_V^0 = 0.07$.}
\label{Fig_random_mass}
\end{figure*}

\subsection{RG equations for model parameters}\label{Sec_full_RGs}

To perform RG analysis, we rescale the frequency and momenta as
follows \cite{Shankar94}
\begin{eqnarray}
\tilde{x}_{\mu} = x_{\mu}b^{-1}.\label{Eq.rescale_x}
\end{eqnarray}
The field operators and model parameters are rescaled in the
following way
\begin{eqnarray}
&&\tilde{\psi} = \sqrt{Z_{\psi}}\psi, \,\,\,\, \tilde{\Delta}_{\Gamma} =
\sqrt{Z_{\Gamma}}\Delta_{\Gamma},\nonumber \\
&& \tilde{v}_{F} = Z_{v}v_{F},\,\,\,\,\tilde{e} =
Z_{e}e\label{Eq.rescale_field}.
\end{eqnarray}
On the basis of the above scaling transformations, we obtain the
complete set of RG equations:
\begin{eqnarray}
\frac{de^2}{d\ln b} &=& e^2 - \frac{N}{8}e^4,\label{Eq.RG_e}\\
\frac{d\Delta_M}{d\ln b} &=& -2\Delta_M(\Delta_M +
\Delta_S-2\Delta_V-4\eta_{\psi}),\label{Eq.RG_M}\\
\frac{d\Delta_S}{d\ln b} &=& 2\Delta_S(\Delta_M +
\Delta_S+2\Delta_V),\label{Eq.RG_V}\\
\frac{d\Delta_V}{d\ln b} &=&0,\label{Eq.RG_A}\\
\frac{dv_{F}}{d\ln b} &=& -(\Delta_M + \Delta_S +
2\Delta_V)v_{F}.\label{Eq.RG_v}
\end{eqnarray}
In the derivation of RG equations, we have redefined the renormalized gauge
coupling as \cite{Kubota2001,Gusynin2001PRD}
\begin{eqnarray}
e^2(p)=\frac{e^2}{1-\Pi(p)},
\end{eqnarray}
which naturally gives rise to the flow Eq. (\ref{Eq.RG_e}).
Moreover, the effective parameter for random potentials are
redefined as
\begin{eqnarray}
\Delta_{\Gamma}/(2\pi v_{F}^2)\rightarrow \Delta_{\Gamma}.\label{Eq.dis_coup}
\end{eqnarray}
Eq.~(\ref{Eq.RG_e}) shows that the flow equation of gauge coupling
is not affected by random potentials at the leading order. This
reflects the fact that random potential does not couple directly to
the gauge boson. Their mutual effects can only be induced by their
separate interaction with Dirac fermions, which are higher order
corrections to the leading order results. Indeed, the flow equation
Eq.~(\ref{Eq.RG_e}) coincides with previous results
\cite{JanssenarXiv, KavehPRB2005, HerbutPRD2016} and exhibits a
stable infrared fixed point at $e_{\ast}^2\sim O(1/N)$ to the
leading order of perturbative expansion. The rest four RG equations,
i.e., Eqs.~(\ref{Eq.RG_M}) - (\ref{Eq.RG_A}) incorporate the influence
of random potentials, and are apparently absent in the clean limit
with $\Delta_M = \Delta_S = \Delta_V = 0$.

According to Eqs.~(\ref{Eq.RG_M}) - (\ref{Eq.RG_A}), we observe that
RM is the only random potential that is directly influenced by the
gauge interaction. For RSP, the flow of $\Delta_S$ depends
sensitively on the interplay of different random potentials. The
effective parameter for RVP, namely $\Delta_V$, simply does not flow
with varying energy scale, which is a consequence of the existence
of a time-independent gauge transformation that ensures RVP
unrenormalized and is valid at any order of loop expansion
\cite{HerbutPRL2008}. Some previous works \cite{HerbutPRL2008,
Vafek08} have studied the RG flow of $\Delta_V$ by considering the
interplay of long-range Coulomb interaction and RVP in graphene. It
was found \cite{HerbutPRL2008, Vafek08} that the parameter
$\Delta_V$ also does not flow.

\section{Interplay between gauge interaction and random potentials}
\label{Sec_single_disorder}

In this section, we analyze the RG solutions and also discuss the
physical effects of random potential on Dirac fermions. Since
$\Delta_V$ does not flow, it can be taken at certain constant. We
will always retain gauge interaction and RVP with $\Delta_V =
\Delta_V^0$ in the system, and study how the system is influenced by
RM and by RSP, respectively. We then consider the most general case
in which the gauge interaction and all three types of random
potentials coexist in the system. Our aim is to find out the
possible infrared fixed points, which will be used to judge the
relevance (or irrelevance) of random potential.

\subsection{Random mass}\label{Sec_random_mass}

In the case of RM, we set $\Delta_S = 0$ and $\Delta_V =
\Delta_V^0$, which simplify the RG equations of $\Delta_M$ and $v_F$
to
\begin{eqnarray}
\frac{d\Delta_M}{dl} &=&
-2\Delta_M(\Delta_M-2\Delta_V^0-4\eta_{\psi}),
\label{Eq.RG_mass_M} \\
\frac{dv_{F}}{dl} &=& -(\Delta_M+2\Delta_V^0) v_{F},
\label{Eq.RG_mass_v}
\end{eqnarray}
where $\eta_{\psi}$ is the anomalous dimension induced by gauge
interaction and $\Delta_V^0$ is a small constant. The solution
for Eq.~(\ref{Eq.RG_mass_M}) has the following form
\begin{eqnarray}
\Delta_{M}(l) = \frac{2\Delta_M^0(2\eta_{\psi}+\Delta_V^0)
e^{4(2\eta_{\psi} + \Delta_V^0)l}}{\Delta_M^0[e^{4(2\eta_{\psi} +
\Delta_V^0)l} - 1] + 2(2\eta_{\psi} +
\Delta_V^0)},\label{Eq_Solu_RG_mass_M}
\end{eqnarray}
where $\Delta_M^0$ is the value of $\Delta_{M}$ defined at the UV
cutoff. It is easy to find that
\begin{eqnarray}
\lim_{l\rightarrow \infty}\Delta_{M}(l) = 2(2\eta_{\psi} +
\Delta_V^0). \label{Eq_Solu_RG_mass_M_lim}
\end{eqnarray}
in the long wavelength limit, which clearly tells us that
$\Delta_{M}^{\ast} = 2(2\eta_{\psi}+\Delta_V^0)$ is the only stable
infrared fixed point. In addition, one can verify that $\Delta_M$
also has an unstable Gaussian fixed point $\Delta_{M}^{\ast} = 0$.
The existence of these two fixed points is illustrated in
Figs.~\ref{Fig_random_mass_a}-\ref{Fig_random_mass_b}. According to
Eq.~(\ref{Eq_Solu_RG_mass_M_lim}), the finite stable infrared point
can be produced by both the gauge interaction and RVP. Therefore, as
along as RM coexists with one of this two kinds of interaction, it
will becomes a marginally relevant perturbation to the system.

To gain a better understanding of the impact of gauge interaction
and RVP on RM, it is interesting to take a look at the coupling
between RM and fermions. When the system contains only RM and Dirac
fermions, RG analysis show that $\Delta_{M}^{\ast} = 0$ is the only
stable fixed point. Although RM is marginally irrelevant in the
absence of gauge interaction and RVP, its importance can be
significantly enhanced by the gauge interaction and RVP. We know
from the above discussion that RM becomes marginally relevant when
it coexists with with the gauge interaction. Such an
interaction-induced enhancement of random potential appears to be a
generic property of several planar strongly correlated systems
\cite{ZhaoPRB2016, Stauber2005PRB, Goswami2011PRL}. Reminding that
we are employing a weak coupling expansion for the coupling between
fermions and random potential. The infrared stable fixed point
generated by gauge interaction is at the order of $O(1/N)$, thus the
weak coupling expansion in the case of RM is reliable. Moreover,
when RM and RVP coexist in the system without gauge interaction, the
finite fixed point is still present. Therefore, RVP can also enhance
RM \cite{OstrovskyPRB2006, FosterPRB2012}.

The enhancement of the role of RM by gauge interaction can be made
clearer by analyzing the low-energy behaviors of fermion velocity
$v_F$. Substituting Eq.~(\ref{Eq_Solu_RG_mass_M}) into
Eq.~(\ref{Eq.RG_mass_v}), and then solving the differential
equations, we obtain
\begin{eqnarray}
v_{F}(l) = \frac{v_{F}^0e^{-2\Delta_V^0
l}}{\sqrt{t_{m}[e^{4(2\eta_{\psi} + \Delta_V^0)l}-1] +
1}},\label{Eq.RM_V_F}
\end{eqnarray}
where
\begin{eqnarray}
t_m \equiv \Delta_{M}^0/2(2\eta_{\psi}+\Delta_V^0)
\end{eqnarray}
and $v_{F}^0$ is the initial value of $v_{F}$ at upper cutoff
$\Lambda$. We first consider the simplest case in which Dirac
fermions couple to RM alone. Since $\eta_{\psi} = \Delta_V^0 = 0$,
the function $v_{F}(l)$ becomes
\begin{eqnarray}
v_{F}(l) = \frac{v_{F}^0}{\sqrt{2\Delta_{M}^0l + 1}}.
\end{eqnarray}
Thus, $v_{F}(l)$ is driven by RM to decrease with growing $l$ grows
and vanishes as $l \rightarrow +\infty$. However, this is not a
exponential decay for which there is no anomalous dimension generated for $v_{F}$.
Indeed, RM only generates a logarithmic correction to $v_{F}$.

\begin{figure}[htbp]
\center \hspace{-4ex}
\subfigure{\includegraphics[width=2.93in]{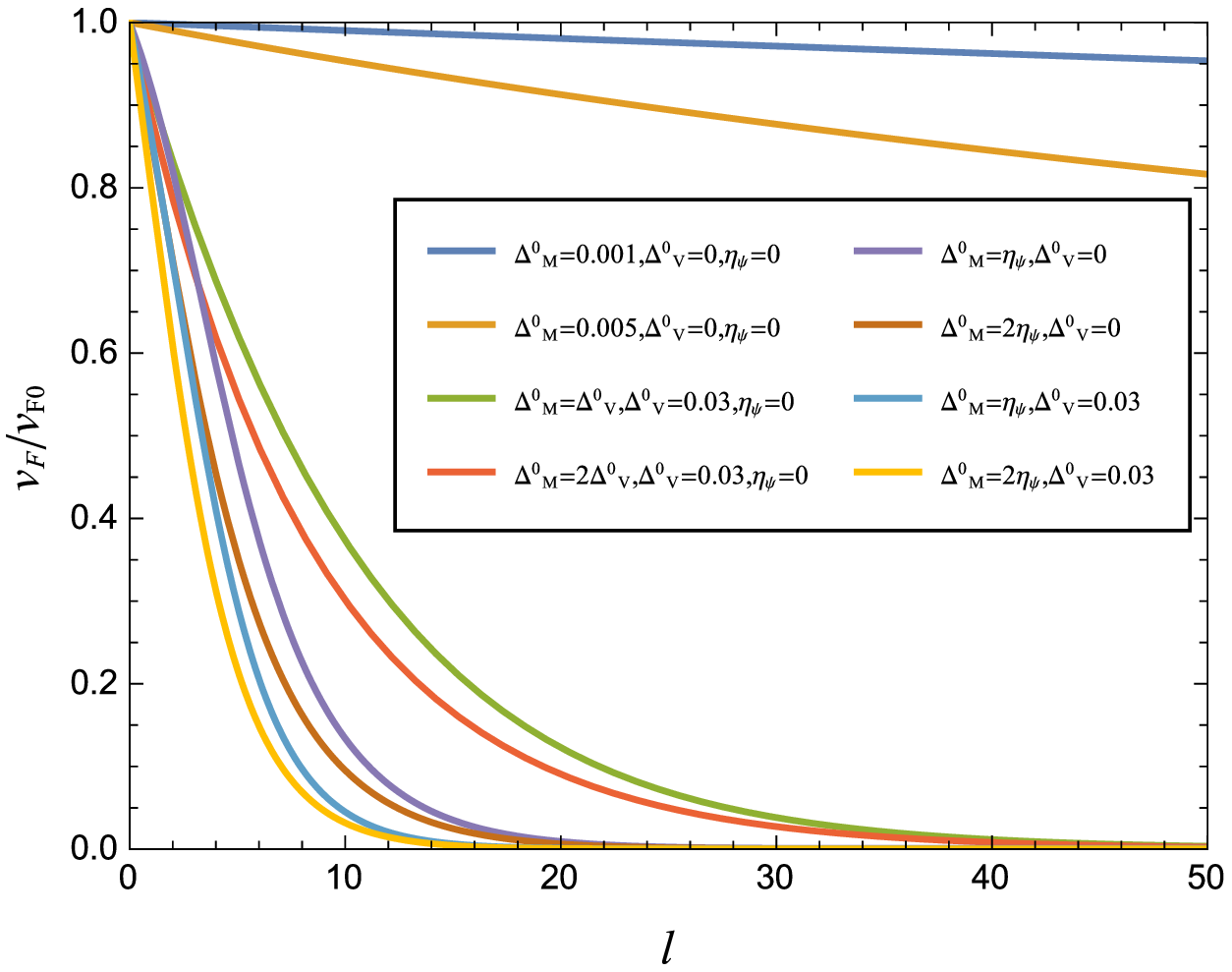}}
\caption{Dependence of $v_{F}(l)$ on the running scale $l$ at
different initial values of $\Delta_M,\Delta_V$. Here, $\eta_{\psi}
= 0$ represents the case without gauge interaction. For nonzero
$\eta_{\psi}$, we assume $N = 4$.}\label{Fig_random_mass_v}
\end{figure}

\begin{figure*}[htbp]
\center \hspace{-4ex}\subfigure{\begin{minipage}{7.8cm}
\includegraphics[width=2.9in]{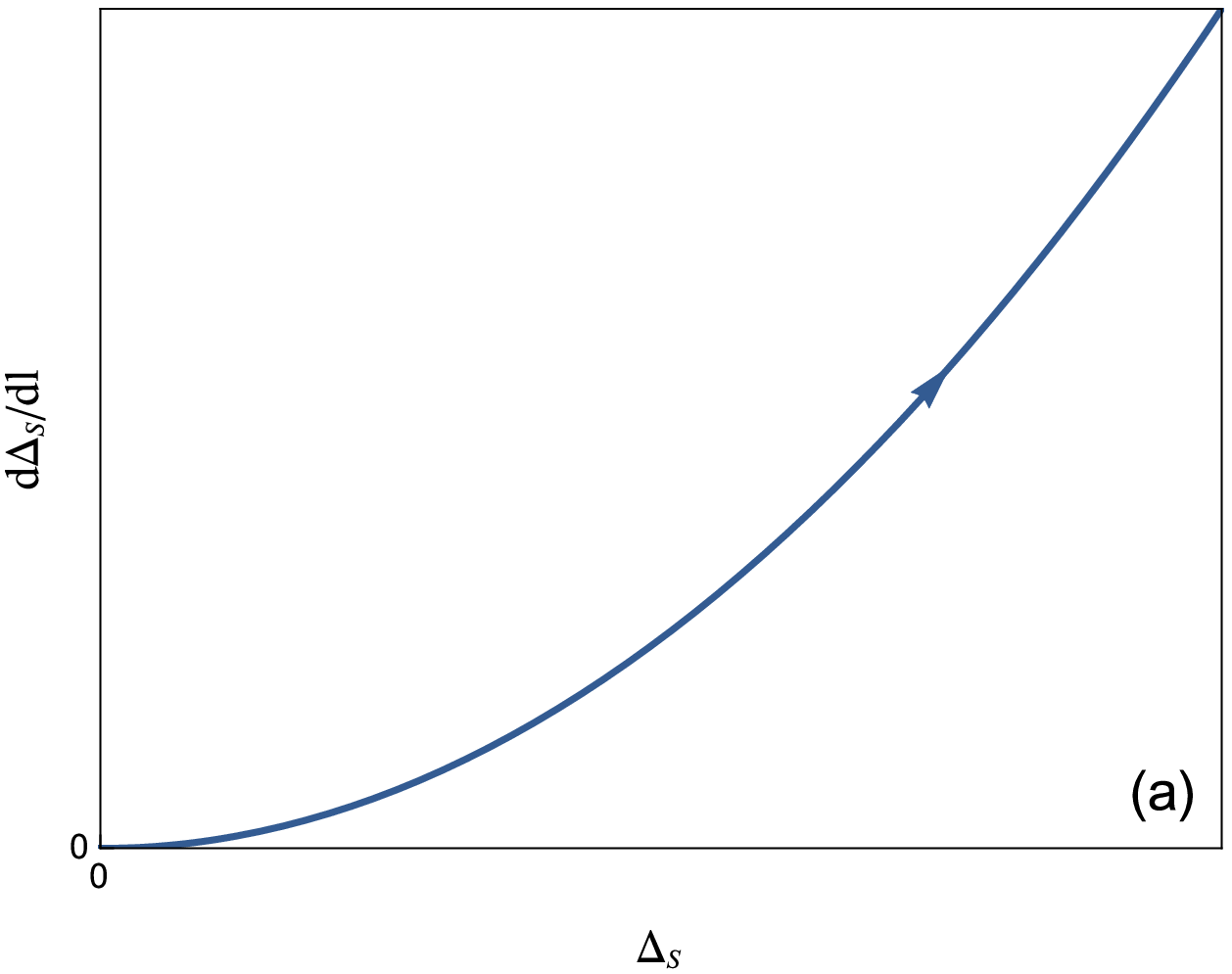}\label{Fig_random_chem_a}
\vspace{-1.3ex}
\end{minipage}}\hspace{2ex}
\subfigure{\begin{minipage}{7.8cm}
\includegraphics[width=2.93in]{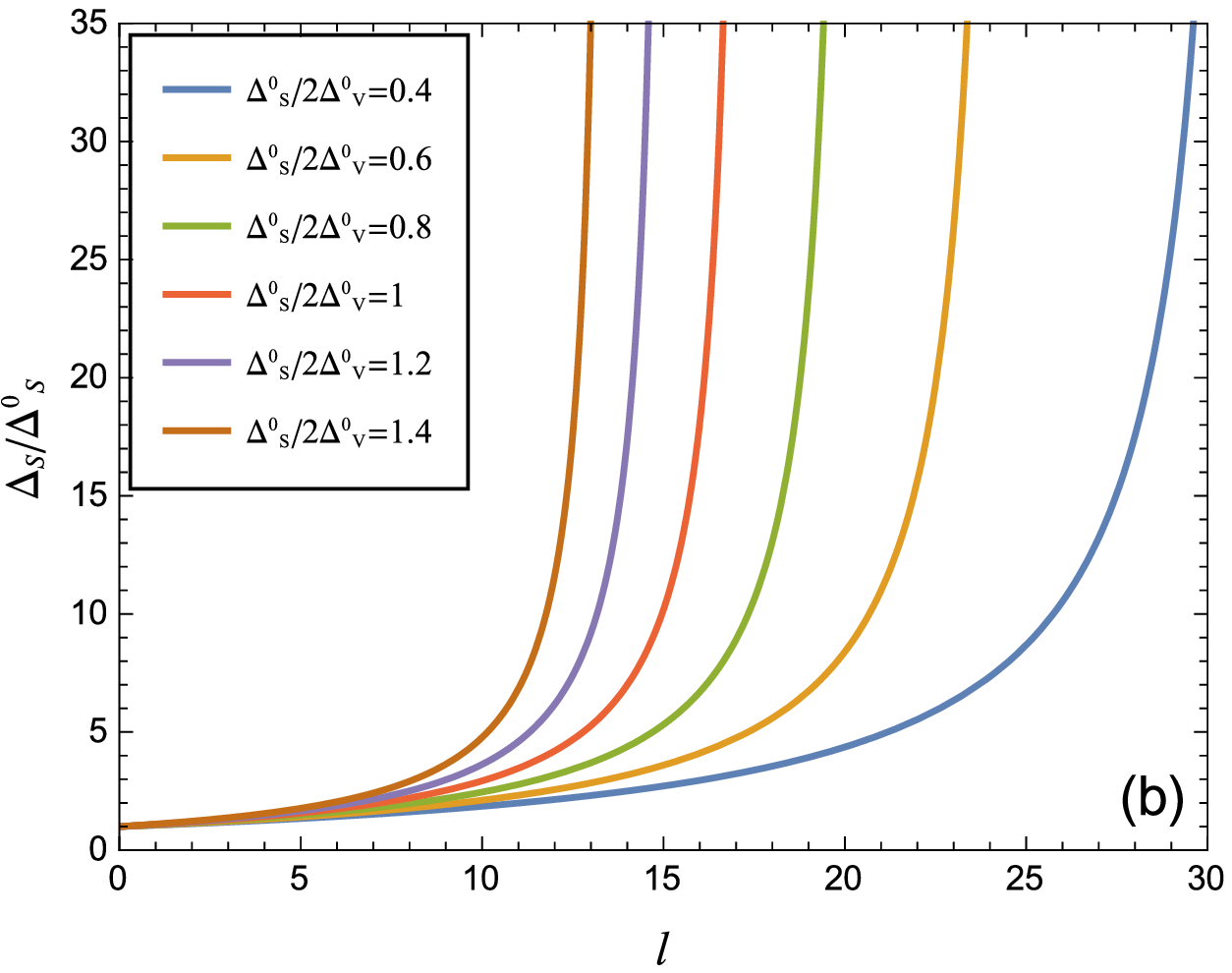}\label{Fig_random_chem_b}
\end{minipage}}
\caption{(a) Flow diagram of $\Delta_S$ for RSP. There is only a
unstable Gaussian fixed point $\Delta_S = 0$. (b) Dependence of
$\Delta_S$ on the running scale $l$ at different initial values of
$\Delta_S/2\Delta_V^0$. Here, $\Delta_V^0 = 0.01$.}
\label{Fig_random_chem}
\end{figure*}

We now add the gauge interaction into the system but keep
$\Delta_V^0 = 0$, and find that Eq. (\ref{Eq.RM_V_F}) becomes
\begin{eqnarray}
v_{F}(l) = \frac{v_{F}^0}{\sqrt{t_{m}^G(e^{8\eta_{\psi}l}-1) + 1}},
\end{eqnarray}
where $t_{m}^G\equiv \Delta_{M}^0/4\eta_{\psi}$. In the
lowest-energy limit, the velocity behaves as
\begin{eqnarray}
v_{F}(l)\big|_{l\rightarrow \infty} \sim v_{F}^0/\sqrt{t_m^G}
e^{-4\eta_{\psi}l}.\label{Eq.RG_mass_v_F_G}
\end{eqnarray}
In this case, $v_{F}(l)$ flows to zero exponentially with growing
$l$. The function Eq.~(\ref{Eq.RG_mass_v_F_G}) can be re-expressed
as a function of momentum $k$ in the form
\begin{eqnarray}
v_{F}(k)\propto k^{\eta_v},
\end{eqnarray}
where $\eta_v = 4\eta_{\psi} = 32/3\pi^2N$ corresponds to the stable
infrared fixed point of RM produced by the gauge interaction. We can
see that $v_F$ now acquires a finite anomalous dimension $\eta_v$,
which takes a universal constant at a given flavor $N$. The
expression of this anomalous dimension is analogous to that obtained
in Ref.~\cite{WangLiu12} which studied the fermion velocity
renormalization in QED$_3$ defined at finite fermion density.
Moreover, this kind of fermion velocity renormalization is a special
property of Dirac fermion systems, including graphene \cite{Vafek07,
Son07, WangLiuNJP12, WangLiu14, Kotov} and high-$T_{c}$
superconductors \cite{Kim97, Xu08, Huh08, Liu12, She15}. It leads to
a series of extraordinary spectral, thermodynamic, and transport
properties of massless Dirac fermions \cite{Vafek07, Son07,
WangLiuNJP12, WangLiu14, Kotov, Kim97, Xu08, Huh08, Liu12, She15}.

We then remove the gauge boson and consider the coexistence of RVP
and RM. In this case, Eq. (\ref{Eq.RM_V_F}) becomes
\begin{eqnarray}
v_{F}(l) = \frac{v_{F}^0e^{-2\Delta_V^0
l}}{\sqrt{t_{m}^d(e^{4\Delta_V^0l}-1) +
1}},\label{Eq.RM_V_F_d}
\end{eqnarray}
where $t_{m}^d\equiv \Delta_{M}^0/2\Delta_V^0$. It is easy to find
that the velocity varies with $l$ as
\begin{eqnarray}
v_{F}(l)\big|_{l\rightarrow \infty} \sim v_{F}^0/\sqrt{t_m^d}
e^{-4\Delta_V^0l},\label{Eq.RG_mass_v_F_d}
\end{eqnarray}
Similarly, $v_{F}(l)$ flows to zero exponentially as $l$ grows. We
then convert Eq.~(\ref{Eq.RG_mass_v_F_d}) to the expression
\begin{eqnarray}
v_{F}(k)\propto k^{\eta_v}.
\end{eqnarray}
with $\eta_v = 4\Delta_V^0$ corresponds to the stable infrared fixed
point induced by RVP and RM. It is thus clear that, similar to the
gauge interaction, RVP can also enhance role played by RM and induce
an anomalous dimension of $v_F$ that is proportional to the strength
of RVP.

When RM coexists with both the gauge interaction and RVP, the
fermion velocity acquires an anomalous dimension $\eta_v =
2(2\eta_{\psi} + \Delta_V^0)$. We present the detailed
$l$-dependence of $v_F$ in Fig.~\ref{Fig_random_mass_v} for the four
different cases discussed in this subsection.

\subsection{Random scalar potential}\label{Sec_random_chem}

We then remove RM and add RSP to the system. By setting $\Delta_{M}
= 0$ and $\Delta_V = \Delta_V^0$, we get the RG equations of
$\Delta_S$ and $v_F$ in the presence of RSP:
\begin{eqnarray}
\frac{d\Delta_S}{dl} &=& 2\Delta_S(\Delta_S+2\Delta_V^0),
\label{Eq.RG_mass_V} \\
\frac{dv_{F}}{dl} &=& -(\Delta_S+2\Delta_V^0) v_{F}.
\label{Eq.RG_chem_v}
\end{eqnarray}
The corresponding flow diagram is schematically shown in
Fig.~\ref{Fig_random_chem_a}-\ref{Fig_random_chem_b}. We find that
there is only one unstable Gaussian fixed point $\Delta_S^{\ast} =
0$. As $l$ increases, $\Delta_S(l)$ exhibits a run-away behavior for
any small initial value. To illustrate this fact more
quantitatively, we obtain the following solution:
\begin{eqnarray}
\Delta_S(l)= \frac{2\Delta_S^0\Delta_V^0}{
(\Delta_S^0 + 2\Delta_V^0) e^{-4\Delta_V^0l}-\Delta_S^0}.
\label{Eq_solu_RG_chem}
\end{eqnarray}
Therefore, the renormalized parameter $\Delta_S$ increases rapidly
with growing $l$ and formally diverges at a finite length scale $l_c
= 1/4\Delta_V^0 \ln \left[1+2\Delta_V^0/ \Delta_S^0\right]$.
However, we should emphasize that $\Delta_S$ does not really
diverge. The superficial runaway behavior of $\Delta_S$ triggers an
instability of the Dirac fermion system, which undergoes a quantum
phase transition that drives the Dirac fermions to move diffusively
\cite{Fradkin1986, Shindou2009, Goswami2011PRL, Roy2014PRB,
Syzranov2015, Kobayashi2014, Biswas2014}. The Dirac fermions acquire
a finite scattering rate $\gamma_{\mathrm{im}}$ in the diffusive
phase, and damp with time in the form \cite{Liliu2010}
\begin{eqnarray}
G(t)\propto e^{-i \gamma_{\mathrm{im}} t}. \nonumber
\end{eqnarray}
It is interesting that such a diffusive transition occurs even if
RSP is arbitrarily weak. According to Eq. (\ref{Eq.RG_mass_V}), we
find that this diffusive behavior already exists when there is only
RM in a Dirac fermion system. It turns out that adding RVP to the
system promotes the role of RM and catalyzes the diffusive
transition.

\begin{figure}[htbp]
\center \hspace{-4ex}
\subfigure{\includegraphics[width=2.9in]{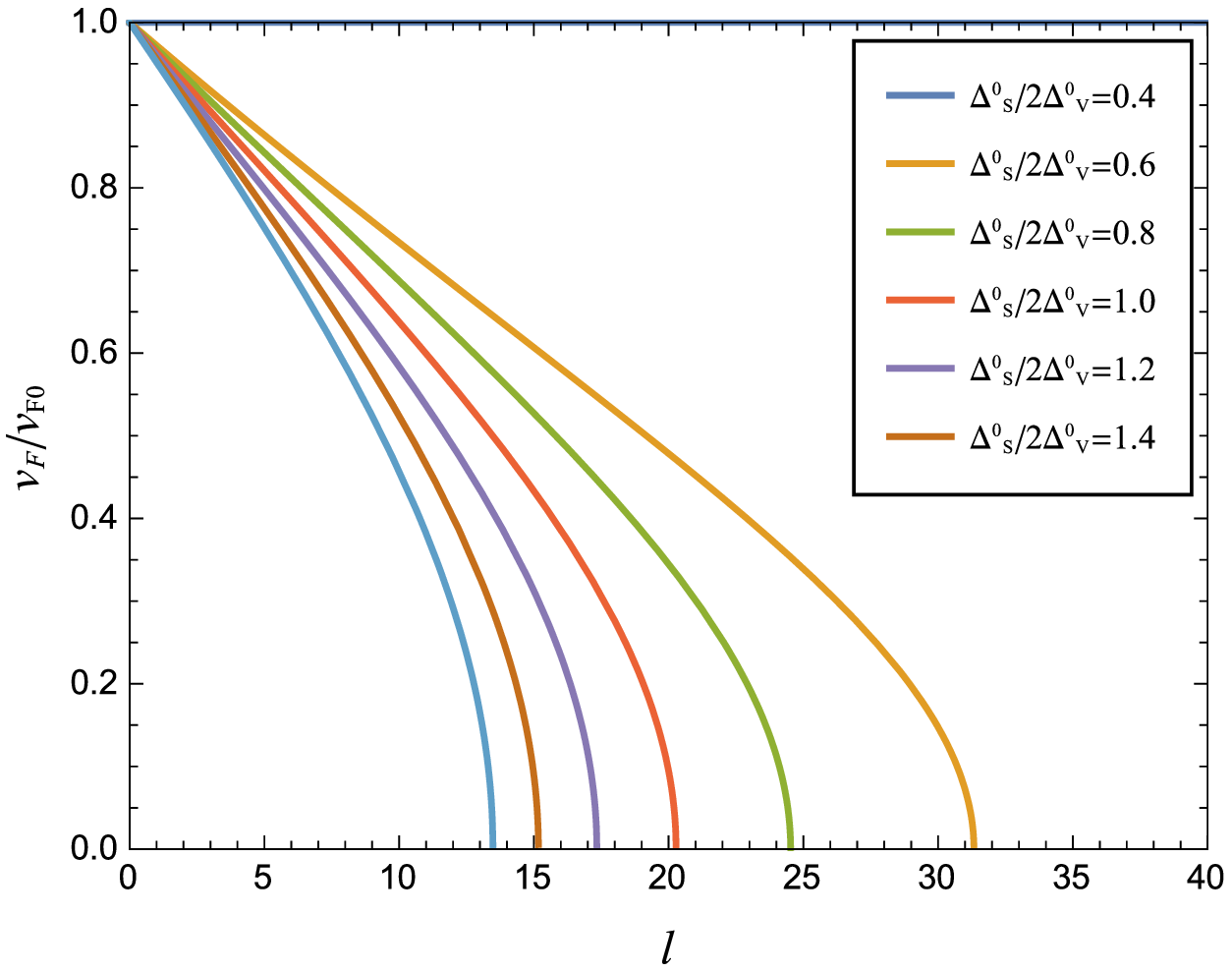}}
\caption{Dependence of $v_{F}(l)$ on the running scale $l$ at
different initial values of $\Delta_S/2\Delta_V^0$. Here,
$\Delta_V^0 = 0.01$.}\label{Fig_random_chem_v}
\end{figure}

The rapid increase of $\Delta_S$ drives $v_{F}$ to vanish at the
length scale $l_c$. To show this, we substitute
Eq.~(\ref{Eq_solu_RG_chem}) into Eq.~(\ref{Eq.RG_chem_v}), and then
get a solution
\begin{eqnarray}
v_{F}(l) = v_{F}^0\sqrt{-t_{s} + e^{-4\Delta_V^0l}(t_s + 1)},
\label{Eq.Chem_vF}
\end{eqnarray}
where $t_s \equiv \Delta_S^0/2\Delta_V^0$. It is easy to verify that
\begin{eqnarray}
v_F(l)\big|_{l\rightarrow l_c} = 0,
\end{eqnarray}
which can also be observed from Fig.~\ref{Fig_random_chem_v}.
However, it is necessary to emphasize that this limiting behavior is
only artificial. The expression of $v_F(l)$ is indeed unreliable
because the perturbative RG method breaks down before $l$ approaches
$l_c$. To obtain a reliable expression for $v_F$ in the low-energy
regime, one should carefully study the diffusive phase
\cite{Roy2014PRB, Moon2014arXiv}, which is very interesting but
beyond the scope of the present work.

\subsection{Random vector potential}\label{Sec_random_gauge}

There exists a peculiar time-independent gauge transformation in the
presence of RVP \cite{HerbutPRL2008, Vafek08}, which renders that
the parameter of RVP is unrenormalized and insusceptible to the
gauge interaction and the rest two types of random potential. Due to
this property, $\Delta_V$ can be regarded as a constant. Now
Eq.~(\ref{Eq.RG_v}) is simplified to
\begin{eqnarray}
\frac{dv_{F}}{dl} = -2\Delta_V^0v_{F},
\end{eqnarray}
which has a solution
\begin{eqnarray}
v_{F}(l) = v_{F}^0 e^{-2\Delta_V^0l},\label{Eq.solu_vF_gauge}
\end{eqnarray}
The velocity depends on $k$ as follows
\begin{eqnarray}
v_{F}(k)\propto k^{\eta_v},
\end{eqnarray}
with $\eta_v = 2\Delta_V^0$.

It is clear that RVP alone is able to induce unusual fermion
velocity renormalization with an anomalous dimension $\eta_v =
2\Delta_V^0$. In particular, the velocity $v_F$ is strongly
suppressed by RVP at low energies, which in turn increases the
effective strength of RM and RSP, as can be seen from
Eq. (\ref{Eq.dis_coup}).

\subsection{Interplay between RM and RSP}\label{Sec_dis_interplay}

We have thus far only considered how RM and RSP are separately
affected by the gauge interaction and RVP. We finally consider the
general case in which both RM and RSP exist in the system, and study
their mutual influence. The flow equations for $\Delta_M$ and
$\Delta_M$ can be written as follows
\begin{eqnarray}
\frac{d\Delta_M}{d\ln b} &=& -2\Delta_M(\Delta_M +
\Delta_S-2\Delta_V-4\eta_{\psi}),\label{Eq.RG_M_inter}\\
\frac{d\Delta_S}{d\ln b} &=& 2\Delta_S(\Delta_M + \Delta_S +
2\Delta_V).\label{Eq.RG_V_inter}
\end{eqnarray}

\begin{figure}[htbp]
\center \hspace{-4ex}
\subfigure{\includegraphics[width=2.9in]{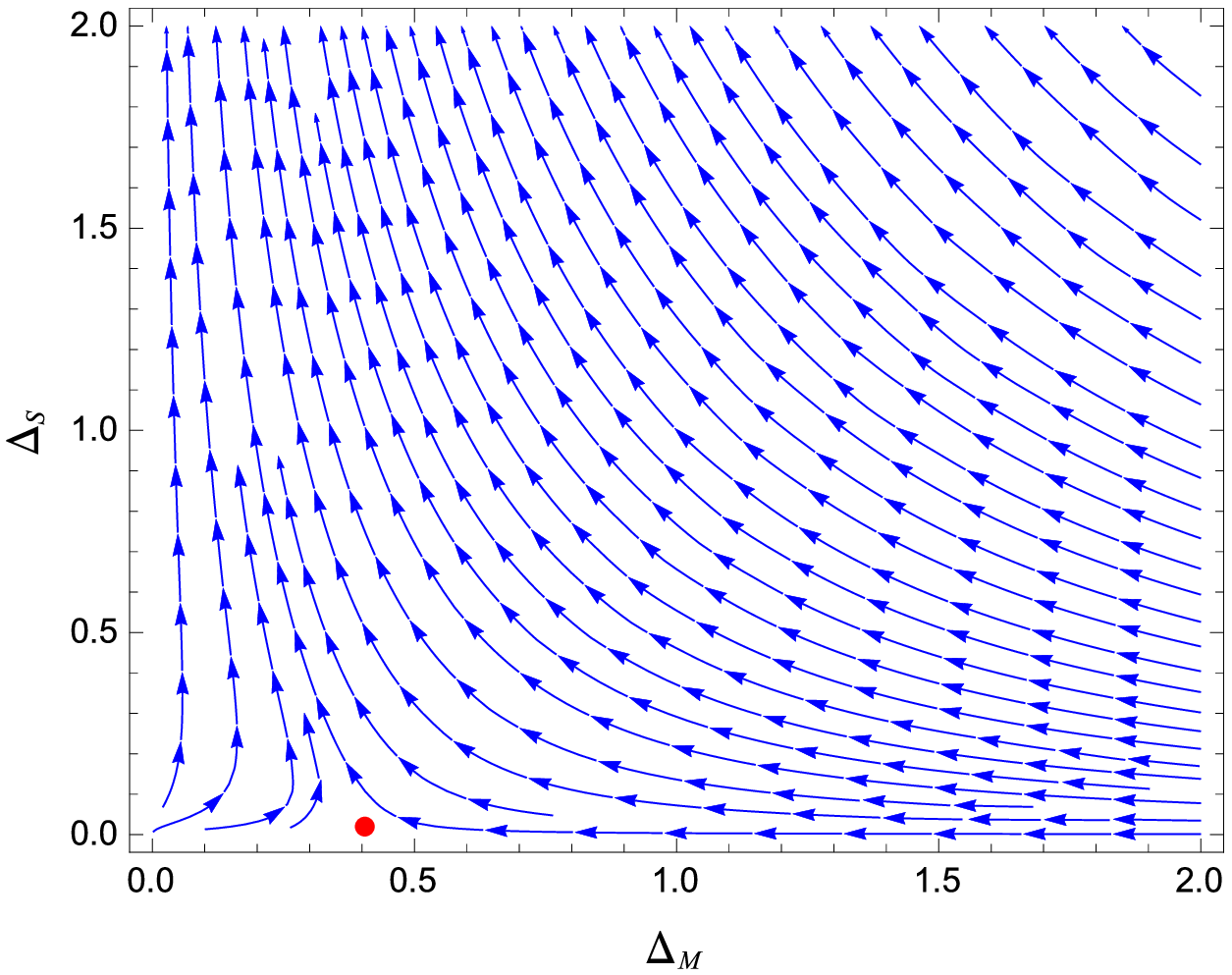}}
\caption{Schematic flow diagram in the plane spanned by $\Delta_M$
and $\Delta_S$, where $N = 4$ and $\Delta_V^0 = 0.07$.}
\label{Fig_flow_diagram}
\end{figure}

The analytical solution of these equations are hard to obtain. We
solve them numerically and present the schematic RG flow diagram in
Fig. \ref{Fig_flow_diagram}. One interesting result is that the
stable infrared fixed point obtained in Sec. \ref{Sec_random_mass}
in the case of RM is eliminated by the coexisting RSP, irrespective
of the strength of RSP. Actually, RM flows to the trivial Gaussian
fixed point as along as RSP is present. It can be concluded that RM
is entirely suppressed by RSP even when the system also contains the
gauge interaction and RVP. Similar to the case without RM, RSP still
shows a run-away behavior and drives a diffusive phase transition.
It turns out that RSP plays a dominant role at low energies and
determines most of the low-energy properties of the system, with RM
and RVP being nearly negligible.

\section{Summary and discussion}\label{Sec_summary}

In summary, we have studied the effects of three types of random
potential on the low-energy behaviors of Dirac fermions in the
context of QED$_3$. After carrying out RG calculations, we have
showed that RM, RSP, and RVP can substantially affect the properties
of mass Dirac fermions. Adding random potentials to the system
explicitly breaks the Lorentz invariance, and leads to fermion
velocity renormalization. We have computed the renormalized velocity
and analyzed its low-energy asymptotic behaviors. The role played by
RM is significantly enhanced by the gauge interaction but RSP and
RVP seems to be insusceptible to the gauge interaction at the
one-loop order. RSP is a marginally relevant perturbation to the
system, and drives the system to undergo a diffusive quantum phase
transition. When three types of random potentials coexist, RSP
dominates and determines the low-energy behavior of the system, with
RM and RVP being nearly ignorable. In the absence of RSP, RVP
promotes RM to become a marginally relevant perturbation, and also
induces an anomalous dimension for fermion velocity.

After determining the infrared fixed point structure of QED$_3$ with
random potentials, the next task could be to analyze the low-energy
behaviors induced by the unusual renormalization of fermion
velocity. It is also interesting to study the rich quantum critical
phenomena at the diffusive quantum critical point, and compute the
associated critical exponents and observable quantities
\cite{Goswami2011PRL, Roy2014PRB, Roy2016PRB, LaiarXiv,
Syzranov2015, Pixley2016, Syzranov2016, Isobe2016PRL}.

In the condensed-matter applications, QED$_3$ may need to be
properly modified. For example, in the effective QED$_3$ theory of
high-$T_c$ superconductors \cite{Lee06, Kim97}, the gauge boson
couples to massless Dirac fermions and additional scalar bosons. It
would be straightforward to include these additional degrees of
freedom into the RG analysis performed in this work.

\emph{Note added.}---After the original version of this paper was
submitted out for publication, we became aware of two related works
by Goswami, \emph{et al.} \cite{Goswami17} and by Thomson and
Sachdev \cite{Thomson17}, who have studied the effects of some sorts
of quench disorder in QED$_3$. For the three sorts of disorder
considered in our paper, our RG equations are in accordance with
those of Ref.\cite{Thomson17}. In particular, the same fixed-point
structure for the strength parameters of gauge interaction and
disorder was obtained in both of our works. Moreover, these three
works all have reached a common conclusion for RMP that there exists
a finite fixed point in the space spanned by the parameters for
gauge interaction and disorder. In the case of RSP, however, the
results obtained in our work and in Ref.\cite{Thomson17} are quite
different from Ref.\cite{Goswami17}, where it is claimed that RSP is
screened and that the fixed point for gauge interaction is stable
against RSP. This difference stems from the fact that both our work
and Ref.\cite{Thomson17} ignore the Feynman diagrams presented in
Fig.1 of Ref.\cite{Goswami17}, which are free of divergence and thus
should not be incorporated in the RG analysis performed in exactly
three space-time dimensions. Moreover, in this paper we have made a
detailed analysis of the interplay between different types of
disorder, which have not been considered in Ref.\cite{Goswami17} and
Ref.\cite{Thomson17}.

\section*{ACKNOWLEDGEMENTS}

P.L.Z. and G.Z.L. would like to thank Jing-Rong Wang for valuable
discussions. The authors acknowledge the financial support by the
National Natural Science Foundation of China under Grants 11274286,
11574285, and 11375168.

\end{document}